\documentclass{aa}
\usepackage{graphicx}
\begin{document}
   \title{Southern Infrared Proper Motion Survey I: Discovery of New
   High Proper Motion Stars From First Full
   Hemisphere Scan}
   \titlerunning{Southern Infrared Proper Motion Survey I}
   \subtitle{}

   \author{N.R. Deacon,
          N. C. Hambly
          \and
          J. A. Cooke
          }
   \authorrunning{Deacon, Hambly \& Cooke}
   \institute{Institute for Astronomy, University of
              Edinburgh, Blackford Hill, Edinburgh EH9 3HJ\\
              \email{nd@roe.ac.uk,n.hambly@roe.ac.uk,j.cooke@roe.ac.uk}
             }

   \date{Received ---; accepted ---}

   \abstract{We present the first results from the Southern Infrared
   Proper Motion Survey. Using 2 Micron All Sky Survey data along with
   that of the SuperCOSMOS sky survey we have been able to produce the
   first widefield infrared proper motion survey. Having targeted the
   survey to identify nearby M, L and T dwarfs we have discovered 72
   such new objects with proper motions greater than 0.5''/yr with 10 of
   these having proper motions in excess of 1''/yr. The most
   interesting of these objects is SIPS1259-4336 a late M dwarf. We
   have calculated a trigonometric parallax for this object of $\pi =
   276 \pm 41$ milliarcseconds yielding a distance of $3.62\pm0.54$pc. We
   have also discovered a common proper motion triple system and an
   object with a common proper motion with LHS 128. The survey
   completeness is limited by the small epoch differences between many
   2MASS and UKI observations. Hence we only recover 22\% of Luyten
   objects with favourable photometry. However the Luyten study is
   itself unquantifiably incomplete. We discuss the prospect of enhancing the survey volume by reducing the lower proper motion limit.
   \keywords{ Astrometry --
                Stars: Low mass, brown dwarfs 
               }
   }

   \maketitle
%
\section{Introduction}
The Solar Neighbourhood provides an opportunity to constrain the field
mass and luminosity functions of Low Mass Stars and Brown Dwarfs. As
Low Mass Stars and Brown Dwarfs represent the products of star
formation they provide clues as to the processes involved. They also
represent a sink for baryonic matter and hence may explain a small
proportion of Galactic dark matter. In addition nearby multiples
can give clues about the formation of such systems.\\ 
Proper motion surveys are a useful tool for discovering nearby
stars. The benchmark for such surveys is Luyten's Half Arcsecond
Catalogue (hereafter LHS) (1979) containing 3587 stars with  $\mu \geq
0.5''/yr$. This survey is incomplete in the southern sky and excludes fields in the galactic plane.
 Modern high proper motion searches seek to complete Luyten's work in the 
 southern sky and to identify objects Luyten failed to find. Some of these use manual 
 blinking techniques such as the Calan-ESO proper motion catalogue (Ruiz et al 
 2001) and the study of Wroblewski and Costa (2001). Recent computational methods 
 include those of Pokorny et al (2003) and Scholz et al
 (2002). Pokorny lists 
 6206 stars with $\mu \geq 0.18''/yr$ found using SuperCOSMOS scans of 131 Schmidt 
 fields. Scholz found 15 stars with high proper motions from APM scans of UKST 
 plates. Eight of these have distances less than 25 pc. Lepine and
Shara (2002) have 
 used POSS plates to search in the northern hemisphere. This has been done with 
 the SUPERBLINK algorithm, in which plates from two different epochs (POSS I 
 and POSS II) are aligned and the later, better POSS II plates degraded to make 
 them of comparable quality to the POSS I plates. The two images are then 
 subtracted making the moving, high proper motion objects
obvious. Teegarden et al's~(2003) recent discovery was made using data from the SkyMorph search for near earth objects, which also picks up stars 
 with a very high proper motion. Most recently Hambly et al. (2004)
have used SuperCOSMOS data to identify five new high proper motion
stars south of $\delta = -60^\circ$.\\

However the majority of these surveys utilise optical $R$ plate data and
many nearby stars are so red that they are barely visible on such
plates. Low Mass Stars and Brown Dwarfs are bright in the near
infrared $I,J,H$ and $K_{S}$. This means that a proper motion search in the
near infrared is ideal for identifying such objects. Near infrared
proper motion surveys have already been used to discover low mass companions to
nearby stars (Scholz et al., 2004) and to confirm that companions are
co-moving with the primary (Seifahrt et al., 2004). However these
are targeted at known nearby stars and are not useful for discovering
new systems. Many proper motion
surveys are also incomplete. SCR1845$-$6357 for example was missed by all surveys prior to
Hambly et al (2004) despite being relatively bright at $R \approx 16$ and
having a high proper motion $\mu > 2''/yr$. It is likely that similar
objects are still to be discovered. In order to identify late M, L and T dwarfs a proper motion
survey in the near infrared is required. For our survey we used
SuperCOSMOS $I$ data and data in $J,H,K_{s}$ from the 2MASS Point
Source Catalogue as our two epochs. 2MASS objects likely to be nearby
low mass stars or brown dwarfs were selected. These were then paired
with likely $I$ plate counterparts. This constitutes the first
widefield infrared proper motion survey ideal for identifying nearby
Low Mass Stars and Brown Dwarfs. Here we present details of the
method used as well as a list of objects found with $\mu>0.5''/yr$.

\section{Candidate Identification}
\subsection{Initial 2MASS object selection}
Firstly candidate 2MASS images had to be selected. To ensure that these
objects were not too close to the photometric limit of the 2MASS
survey, the objects had to be brighter than $J=16$. Elliptical objects may
have inaccurate photometry and astrometry, hence selected objects had
to  have an axial ratio less than 1.4. Crowded regions near the
galactic plane will produce many spurious detections of high proper
motion objects. In order to reduce this crowding any object with
$|b|<15^{\circ}$ was excluded. The 2MASS Point Source Catalogue includes a
parameter indicating the proximity of the nearest source; the
Executive Summary for the catalogue states that any object closer
than 6'' to its nearest neighbouring source must be treated with
caution. Hence such objects were excluded. Objects categorised by the
2MASS survey as being associated with extended
sources or minor planets were also removed.\\

Objects which met these criteria were then subjected to a series of
colour cuts.\\

\begin{figure}
   \resizebox{\hsize}{!}{\includegraphics{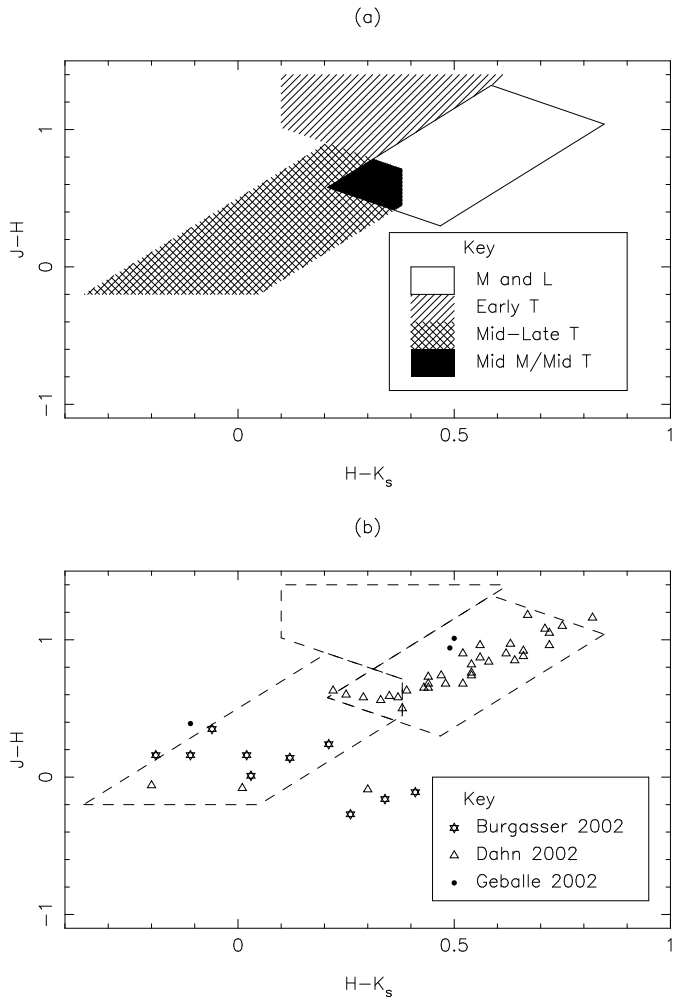}}
      \caption{The regions in the colour-colour diagram within which
   objects were deemed to be M, L or T dwarfs.}
         \label{ccut}
   \end{figure}

In order to make the process as efficient as possible, only objects
whose photometry suggested that they were M, L and T dwarfs were
selected.  In order to do this regions on a colour-colour diagram were
marked out as likely to contain objects of a particular spectral
type. Figure~\ref{ccut}(a) shows these regions and
Figure~\ref{ccut}(b) shows the colours of objects taken from various
surveys. Note that because there is an
overlap between the mid M and mid T spectral classes on a
colour-colour diagram any object that falls in this region is treated
as if it could be either spectral type. Several early M dwarfs fall
outside the bounds of the M dwarf region. When the full survey is
produced this will have to be included in completeness estimates. Many
T dwarfs fall outside the T dwarf region, this scatter is due to
photometric errors. As we have set an $I$ plate limit and an $I-J$ cut
we will only
identify the brightest T dwarfs. Dimmer objects have higher
photometric errors and hence higher scatter. As these objects are
already excluded due to our other photometric cuts it does not matter
if they fall outside the T dwarf region on the colour colour diagram.\\ 

\begin{figure}
   \resizebox{\hsize}{!}{\includegraphics{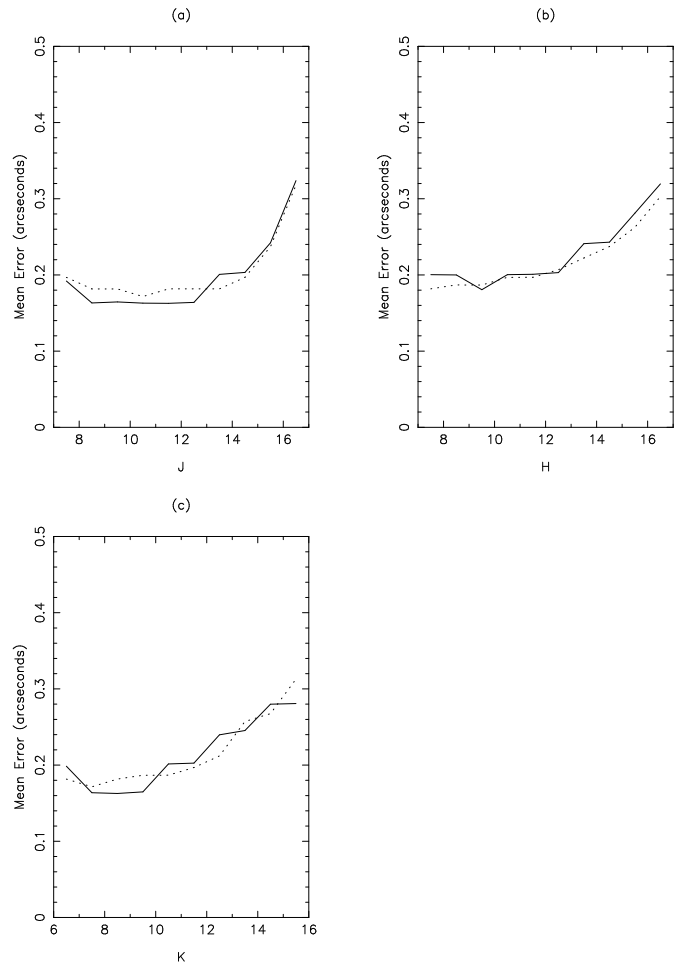}}
      \caption{The relative positional errors between SuperCOSMOS and
2MASS. The solid line represents the errors in Right Ascension  while
the dotted line shows those in Declination.}
         \label{2MASSerrplot}
   \end{figure}

Once objects have passed the colour cuts and been categorised it is
then necessary to find if they are in fact moving with significant
proper motions. Figure~\ref{2MASSerrplot} shows the relative
positional errors between SuperCOSMOS and 2MASS data. These were calculated by finding
the positional shift of a set of 2MASS objects and calculating the
mean error using a method of median absolute deviation. On the basis of these results it
was decided to set one arcsecond (roughly 4-5 $\sigma$) as the lower
``movement cut''. This ensures that only objects which appear to have
moved from one epoch to the other are included in the sample. Any 2MASS object found to have an $I$ plate
counterpart within one arcsecond was discounted as not having a
significant enough proper motion. This cut has implications for proper motion completeness which will have to be taken into account. Also to remove spurious detections
due to bright stellar halos, any 2MASS object within 10'' of an $I$
plate image flagged as being close to a bright star was
removed. Images blended on the $I$ plates can cause positional
offsets, hence any 2MASS object within 10'' of a bright ($I<14$) highly
elliptical object or 6'' of one deblended by SuperCOSMOS software was
removed. 
\subsection{Pairing with potential $I$ plate counterparts} 
Once it is established that objects have appeared to move their $I$
plate counterparts need to be identified. In order to do this a maximum
proper motion of $\mu_{max} = 10''/yr$ was set meaning only $I$ plate objects
within a maximum radius $r_{max}= \mu_{max} \Delta t$ of the 2MASS object would
be paired, where $\Delta t$ is the time elapsed between the two observations. This maximum proper motion was chosen to be roughly equal
to that of Barnard's star, the star with the highest known proper
motion, this allows us to identify very high proper motion stars
without being swamped by spurious pairings. Once an $I$ plate object was identified as within $r_{max}$ of
the target it had to pass a series of tests to prove it was
indeed a plausible counterpart to the 2MASS object.\\
Firstly the $I$ plate object had to have astrometry suggesting it was a high proper
motion star. These data come from the calculated astrometric solutions
in the SuperCOSMOS sky survey (Hambly et al. 2001). Any $I$ plate object
with an astrometric solution that places it at the same position (within the
proper motion errors quoted in the astrometric solution) as the 2MASS
object at the epoch of the 2MASS observation will be included as a
possible counterpart. Many objects will either be so red that they will only appear on the $I$
plates also objects moving
with high proper motions will have moved too far to be paired in other bands so will be recorded as only being on the $I$ plates. Hence any object appearing only on the $I$ plates was included
as a possible counterpart. Finally some $I$ plate objects will be incorrectly
paired with unrelated objects on the $B_{J}$ or $R$ plates and hence
will have an inaccurate astrometric solution. Such
objects will have a high value of $\chi_{\nu}^{2}$ ($\chi^{2}$ per
degree of freedom), hence any object
with  $\chi_{\nu}^{2}>2.0$ was included as a possible counterpart.\\
The $I$ plate images also had to pass tests to indicate that they were
good stellar images. Objects deblended by the SuperCOSMOS software were excluded as it is difficult
to gather accurate astrometric information from them. Images with
axial ratios higher than 1.7 were also excluded as they may have poor
astrometry or may be blended objects which have not been separated by
the software. $I$ plate counterpart objects also had to be stellar in
nature and not be in close proximity to bright stars.\\
Candidates for $I$ plate counterpart images also had to pass a
series of photometric tests. Firstly to ensure that the $I$ plate
counterparts were red enough to be true Low Mass Stars or Brown Dwarfs
any object which had a UKST $R$
plate magnitude had to pass the cut $(R-I)>2$. As noted in Section 2.1, 2MASS objects were classified as being, M
and L dwarfs, early T dwarfs, mid-late T Dwarfs or as lying in the
overlap region on the colour-colour diagram between M dwarfs and mid-T
Dwarfs. Objects classified as M and L dwarfs had to fulfill the condition
$1.0<I-J<4.8$. Objects classified as early T Dwarfs or mid-late T
dwarfs had to be within the range $4.0<I-J<5.5$. Any object falling
in the overlap region between M dwarfs and mid-T dwarfs could be of
either type so had to conform to the conservative cut of
$1.0<I-J<5.5$\\
In order to ensure that the $I$ plate images selected were true
counterparts and not unrelated objects two additional processes were
carried out. Firstly it was checked that the $I$ plate image was not
associated with (within one arcsecond of) another 2MASS
image. Secondly inspection by eye of the images ensured that only true
high proper motion objects were selected.      

\section{Determination of Proper Motions}
Objects in this paper are initially selected using proper
motions calculated by finding the distance an object had moved between
the two epochs and dividing by the time elapsed. Any object with such
a calculated proper motion greater than 0.4''/yr was selected to have
its proper motion more accurately calculated. The SuperCOSMOS Sky
Survey software calculates proper motions using $B_J,I$ and two epoch
$R$ measurements for each object in the
database. Hence, in theory, we should have access to astrometric solutions for all the
objects selected so far. Unfortunately many do not have suitable
solutions, stars with very high proper motions will have large shifts
in their positions from plate to plate meaning they will often be too far
apart from one epoch to the next for the the SSS software to
identify them as the same object. Hence such stars will not have
calculated proper motions. Some high proper motion objects may be
spuriously paired with faint unrelated objects on one or more plates,
producing an unreliable astrometric solution. Also the reddest objects
may only appear on the $I$ plate and hence will have no calculated
astrometric solution. In order to make sure
that all objects had a reliable astrometric solution a procedure to
calculate the astrometric solution using only $I$ plate and 2MASS data
has been developed. Firstly a set of reference stars with $I$ and $J$
magnitudes similar to the star whose proper motion we wish to
calculate was selected. These reference stars are then used to produce a linear fit
between the plates, to account for plate to plate scaling and
orientation errors. The stars positions are then corrected for plate to
plate errors and used to find the proper motion. The plate to plate
fit also provides an estimate of the typical error for objects near
the target star and this is used to calculate the error in the proper
motion. In the case where an object has a calculated astrometric
solution both from the $I$ plate-2MASS fitting procedure and from the
SuperCOSMOS software the most accurate one is quoted. Unfortunately
some objects did not have enough reference stars to produce a plate
model. These had their proper motions found in a same way to those
with enough reference stars with the only difference being the lack
of corrections for plate to plate errors. The error estimate for these
stars was calculated using the positional errors shown in Figure~\ref{2MASSerrplot}.  
\section{Results}
\begin{figure*}[htb]     
        \begin{center}
   \resizebox{\hsize}{!}{\includegraphics{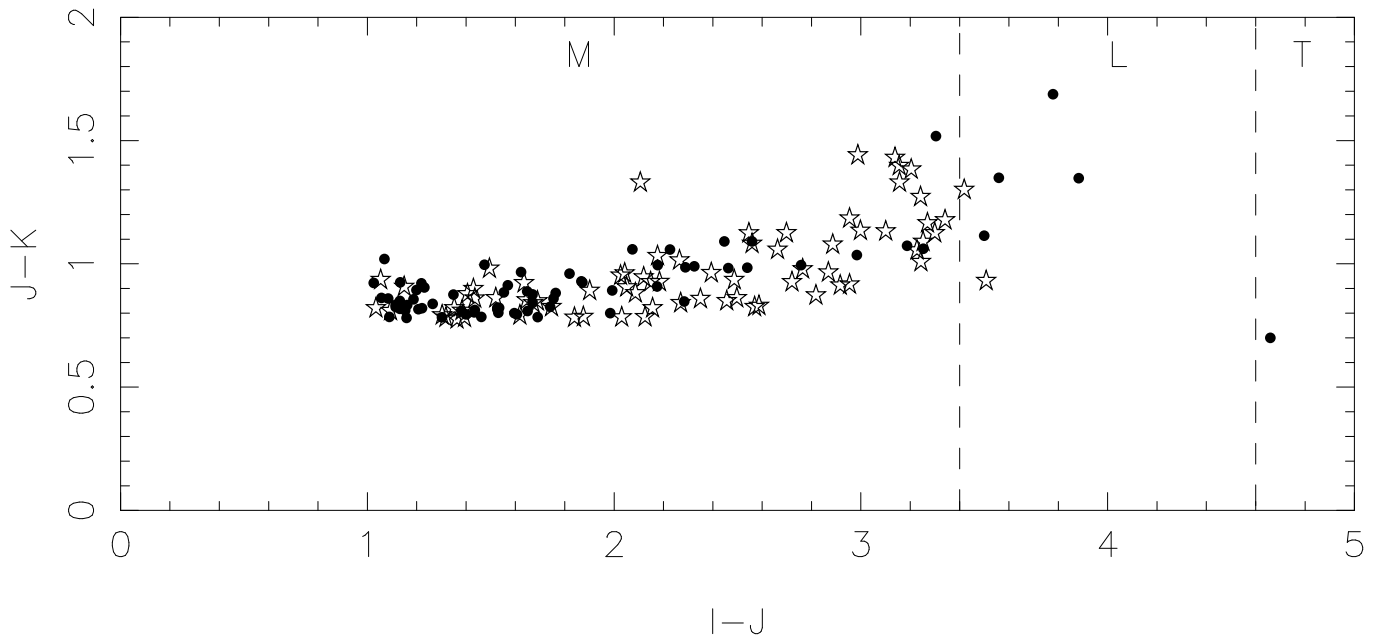}}
\end{center}
\caption{A two colour diagram for all the objects found in the survey. The star symbols represent new discoveries while the solid circles are previously known objects}
         \label{2colour}
\end{figure*}

\begin{figure*}[htb]     
        \begin{center}
\resizebox{\hsize}{!}{\includegraphics{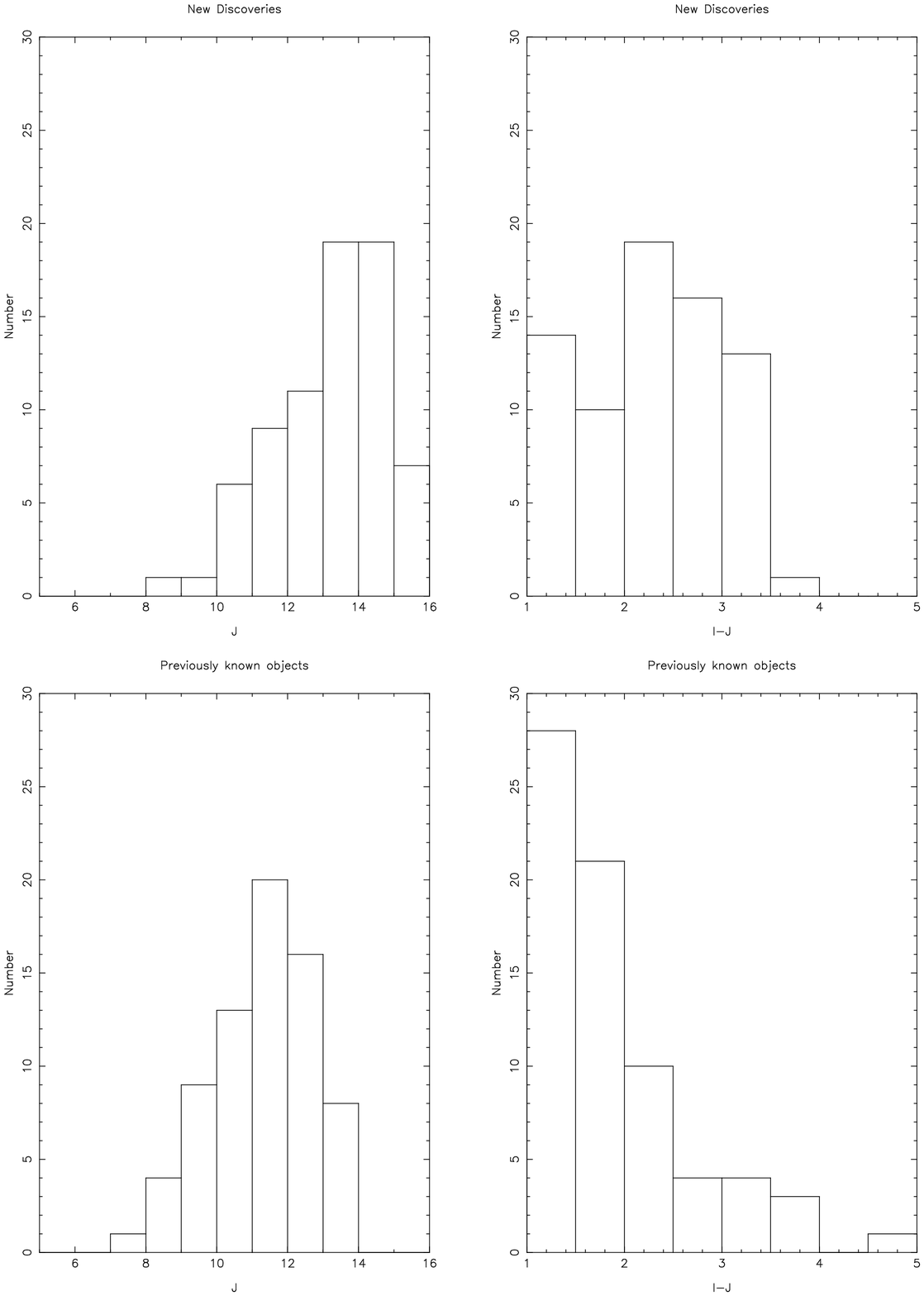}}
\end{center}
\caption{Histograms showing the magnitude ($J$) and colour ($I-J$) of
both the newly discovered and previously discovered objects.}
         \label{phothist}
\end{figure*}
Appendix~\ref{Tabs} contains the astrometric data for all
the objects identified with proper motions greater than
0.5''/yr. There are 73 new objects and 71 objects identified by previous
surveys. The $I, J, H$ and $K_s$ photometry for each object are also
listed. Figure~\ref{2colour} shows the colours of the newly discovered
objects. The reddest newly discovered object is SIPS0641$-$4322 which
has colours suggesting it is an early L dwarf. Objects such as this
would be difficult to find using purely optical proper motion
surveys. Figure~\ref{phothist} shows histogram of the number of newly
discovered and previously known objects vs. magnitude ($J$) and colour
($I_J$). It is clear that the newly discovered objects are generally
fainter and redder than those previously known. Below we present notes on the most interesting individual
newly identified objects identified.   
\subsection{Notes On Individual Objects}
\subsubsection{SIPS1259$-$4336}
Infrared colours indicate this object is an M8-9
dwarf~\cite{Kirkpatrick}. It is one magnitude fainter than the similar
object SCR1845$-$6357~\cite{Hambly2003} placing it at a distance of
roughly 6pc. In addition its high proper motion suggests that it is a
nearby star. Hence as there were
several non-survey plates available for this object in the UKST plate
library a trigonometric parallax measurement was carried out. This
draws on techniques used in Deacon \& Hambly (2001) and Deacon et
al. (2005). Four of the non-survey plates were selected for their good parallax factors
and were scanned on SuperCOSMOS and used along with the existing UKST R
and ESO R plates. Details of the plates used are shown in
Table~\ref{1259plates}. 

\begin{figure}[htb]     
        \begin{center}
   \resizebox{\hsize}{!}{\includegraphics{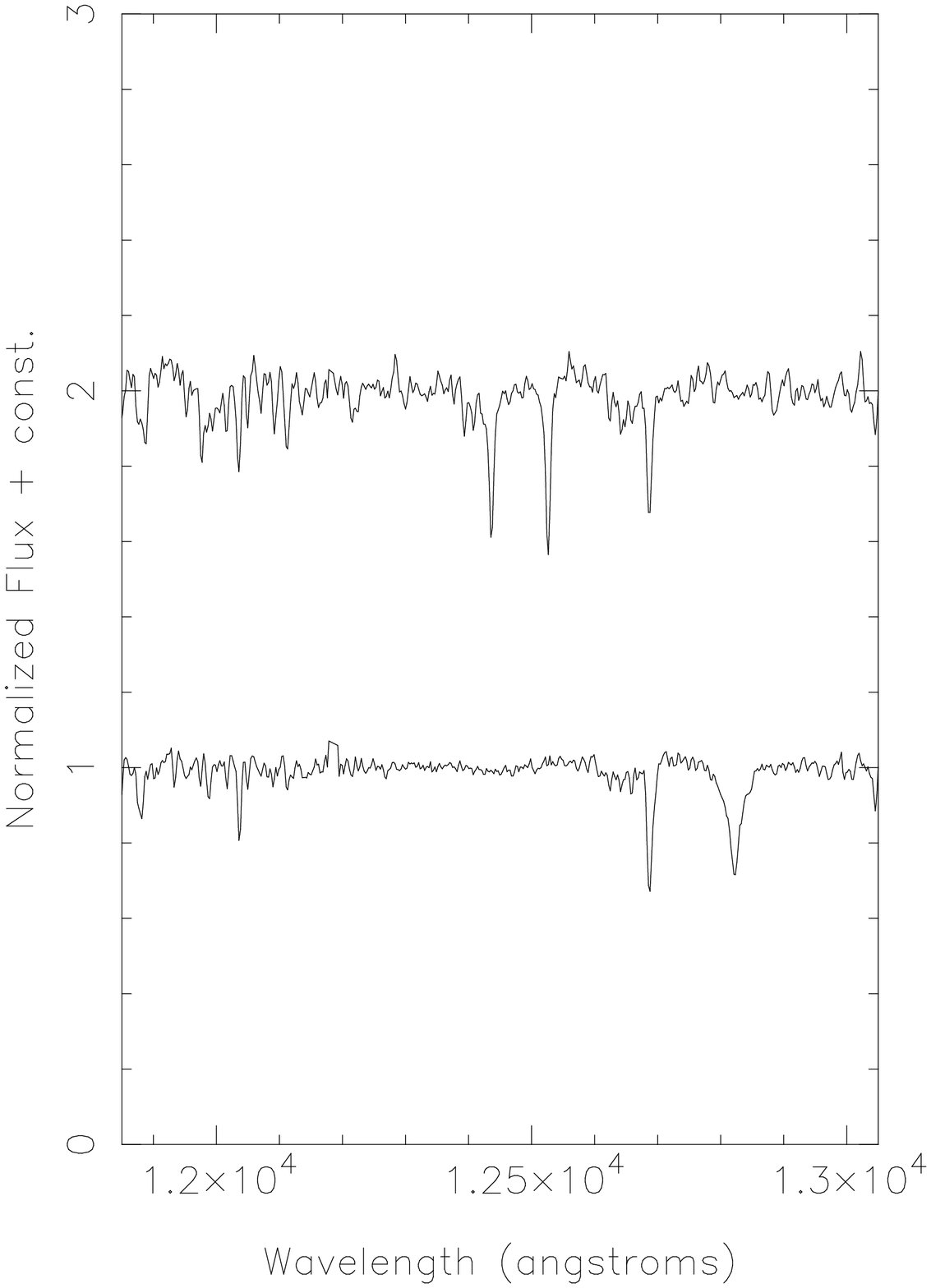}}
\end{center}
\caption{The normalized spectrum of SIPS1259$-$4336 (top) along with that of an A0  standard star. The potassium doublet at 1.25 microns can clearly be seen.}
         \label{JLspec}
\end{figure}

\begin{figure}[htb]     
        \begin{center}
   \resizebox{\hsize}{!}{\includegraphics{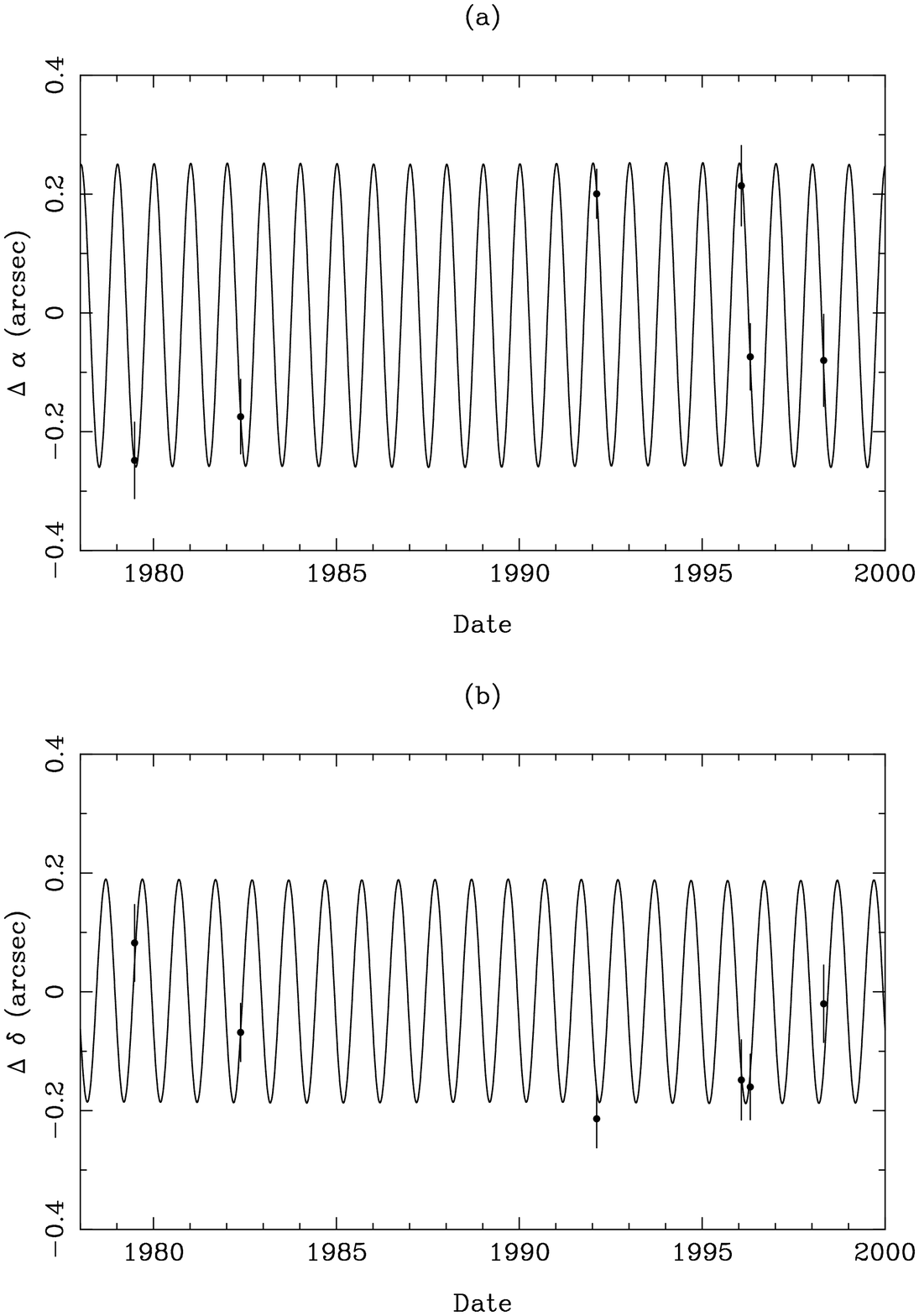}}
\end{center}
\caption{The deviation from the proper motion in RA and Dec plotted against
         time. The line shown is the path of parallax motion predicted
         by the astrometric solution}
         \label{1259parallaxa}
\end{figure}

\begin{figure}[htb]     
        \begin{center}
\resizebox{\hsize}{!}{\includegraphics{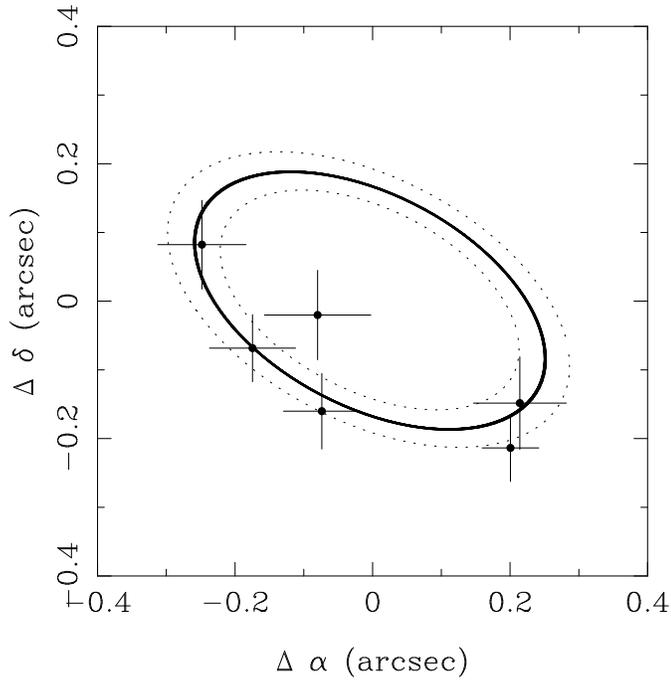}}
\end{center}
\caption{The parallax ellipse traced out by SIPS1259-4336; the dotted
   lines represent one sigma upper and lower limits on the parallax.}
         \label{1259parallaxb}
\end{figure}

\begin{figure}[htb]     
        \begin{center}
\resizebox{\hsize}{!}{\includegraphics{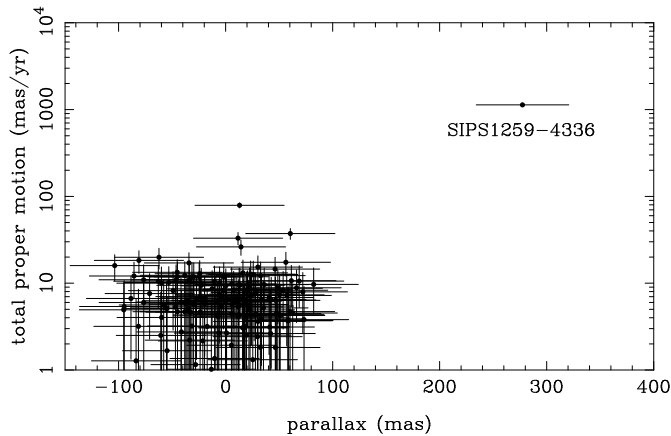}}
\end{center}
\caption{Comparison of the fitted parallax and proper motions for
   the reference stars with SIPS1259-4336.}
         \label{1259paraplot}
\end{figure}

 \begin{table*}[h!!]
      \caption[]{Schmidt photographs used in this study; relative
   astrometric quality is indicated (see text). One plate was excluded due to poor astrometric quality.}
         \label{1259plates}
\begin{tabular}{cccccccccl}
\hline
Plate & Date     & LST &Zenith& Emul- & Filter & Exp.  & $\sigma_x$ & $\sigma_y$ &Material\\
No.   & (dd/mm/yy) &     &Angle& sion  &        & (min) & (mas)      & (mas)      &\\
\hline
\multicolumn{10}{c}{Plates Used}\\
\hline
R 5108&25/06/79&1254&$12.5^{\circ}$&IIIaF&RG 630&800&64&65&Original
non-survey \\
&&&&&&&&&plate\\
ESOR4809&19/05/82&1200&$18.6^{\circ}$&IIIaF&RG630&120&63&49&Copy of ESO
survey \\
&&&&&&&&&plate\\
OR14770&12/02/92&1245&$12.6^{\circ}$&IIIaF&OG 590&600&41&49&Original
survey \\
&&&&&&&&&plate\\
OR16928&26/01/96&1142&$19.6^{\circ}$&4415&OG 590&150&68&67&Original
non-survey \\
&&&&&&&&&film\\
OR17069&26/04/96&1322&$13.1^{\circ}$&IIIaF&OG 590&100&56&55&Original
non-survey \\
&&&&&&&&&plate\\
OR17997&28/04/98&1314&$12.7^{\circ}$&IIIaF&OG 590&600&77&65&Original
non-survey \\
&&&&&&&&&plate\\
\hline
\end{tabular}
   \end{table*}
The data from these plates were reduced using the same method as
SCR1845-6357. A total of 148 reference stars were used to produce the
plate to plate models. Once these had been applied to correct for
plate to plate errors the astrometric solution could be found, this is
shown in Table~\ref{1259Res}. The deviation from the proper motion is
shown in Figure~\ref{1259parallaxa} with the parallax ellipse traced
out by the object shown in Figure~\ref{1259parallaxb}. To check for
any systematic errors the reference stars were run through the fitting
proceedure. Their proper motions and parallaxes are show in
Figure~\ref{1259paraplot} along with those for SIPS1259-4336. It is
clear that there is no significant offset in the parallax and that
SIPS1259-4336 is well separated from the mass of reference stars. The
mean $\chi_{\nu}^{2}$ for the reference stars was found to be 0.97
indicating good model fits.
\begin{table}
        \begin{center}
      \caption[]{The full astrometric solution for SIPS1259-4336}
         \label{1259Res}
\begin{tabular}{cccl}
\hline
Parameter&Fitted Value&Error&Units\\
\hline
RA on 01/01/2000 & $12^{\rm h}59^{\rm m}04^{\rm s}\!.760$ & 0.047~as& --- \\
Dec on 01/01/2000& $-43^{\circ}36'24"\!.21$ & 0.047~as& --- \\
$\mu_{\alpha}$&1.105&0.004&as/yr\\
$\mu_{\delta}$&-0.262&0.004&as/yr\\
$\pi$&276&41&mas\\
$\chi_{\nu}^{2}$&0.65&&\\
\hline
\end{tabular}
        \end{center}
   \end{table}
Spectroscopic data for this object were obtained from IRIS2 at the Anglo-Australian Telescope. These data were reduced and the spectrum in the $J_l$ band is shown in Figure~\ref{JLspec}. Using the equivalent width of the potassium doublet as a diagnostic (Gorlova et al., 2003) we find that it is likely to be an M8 dwarf.
The calculated parallax of SIPS1259-4336 of $276 \pm 41$
milliarcseconds relates to a distance of $3.62\pm0.54$ parsecs. This
is nearer than expected as while the object is of similar colour to
SCR1845-6357 it is one magnitude dimmer, hence it should be much
further away. However if SCR1845-6357 is an unresolved double then the
distance quoted here could be accurate.
\subsubsection{SIPS1910$-$4132}
\begin{table*}
   \caption[]{}
         \label{1910}
\begin{tabular}{lcccccc}
\hline
Name & Position (J2000)&Epoch& $\mu_{tot}$ & PA & $\sigma_{\mu}$\\
\hline
SIPS1910$-$4133A&19 10 34.50 -41 33 32.5&1983.346&0.742&174.4&0.017\\
SIPS1910$-$4133B&19 10 45.87 -41 33 28.9&1983.346&0.748&173.0&0.021\\
SIPS1910$-$4132C&19 10 33.49 -41 32 38.8&1983.346&0.738&174.7&0.016\\
 \hline
\end{tabular}
\end{table*}
\begin{table*}
   \caption[]{}
         \label{1910phot}
\begin{tabular}{lcccc}
\hline
Object & $I$ & $J$ & $H$ & $K_{S}$\\
\hline
SIPS1910$-$4133A&11.008&9.851&9.245&9.032\\
SIPS1910$-$4133B&11.738&10.61&10.002&9.739\\
SIPS1910$-$4132C&12.577&11.147&10.552&10.249\\
 \hline
\end{tabular}
\end{table*}
When the candidate image for SIPS1910$-$4232 was inspected it was found
that there were two objects moving with a similar proper
motion. Astrometric solutions for these companion objects were found
by the $I$ plate-2MASS method as described in Section 3 and are shown in Table~\ref{1910}. It is
clear that the three objects share a common proper motion and hence
are likely to be a triple system. Table~\ref{1910phot} shows the
photometry for these objects. They all appear to be mid-late M dwarfs.
\subsubsection{SIPS0251$-$0352}
This object has the highest proper motion of the newly discovered
objects. Its colours indicate that it is an early L dwarf (Kirkpatrick
et al. 2000). Comparing its $J$ magnitude with mean
characteristics of L dwarfs in Vbra et al. (2004) shows that it
appears to be within 20pc.
    \begin{figure}
   \resizebox{\hsize}{!}{\includegraphics{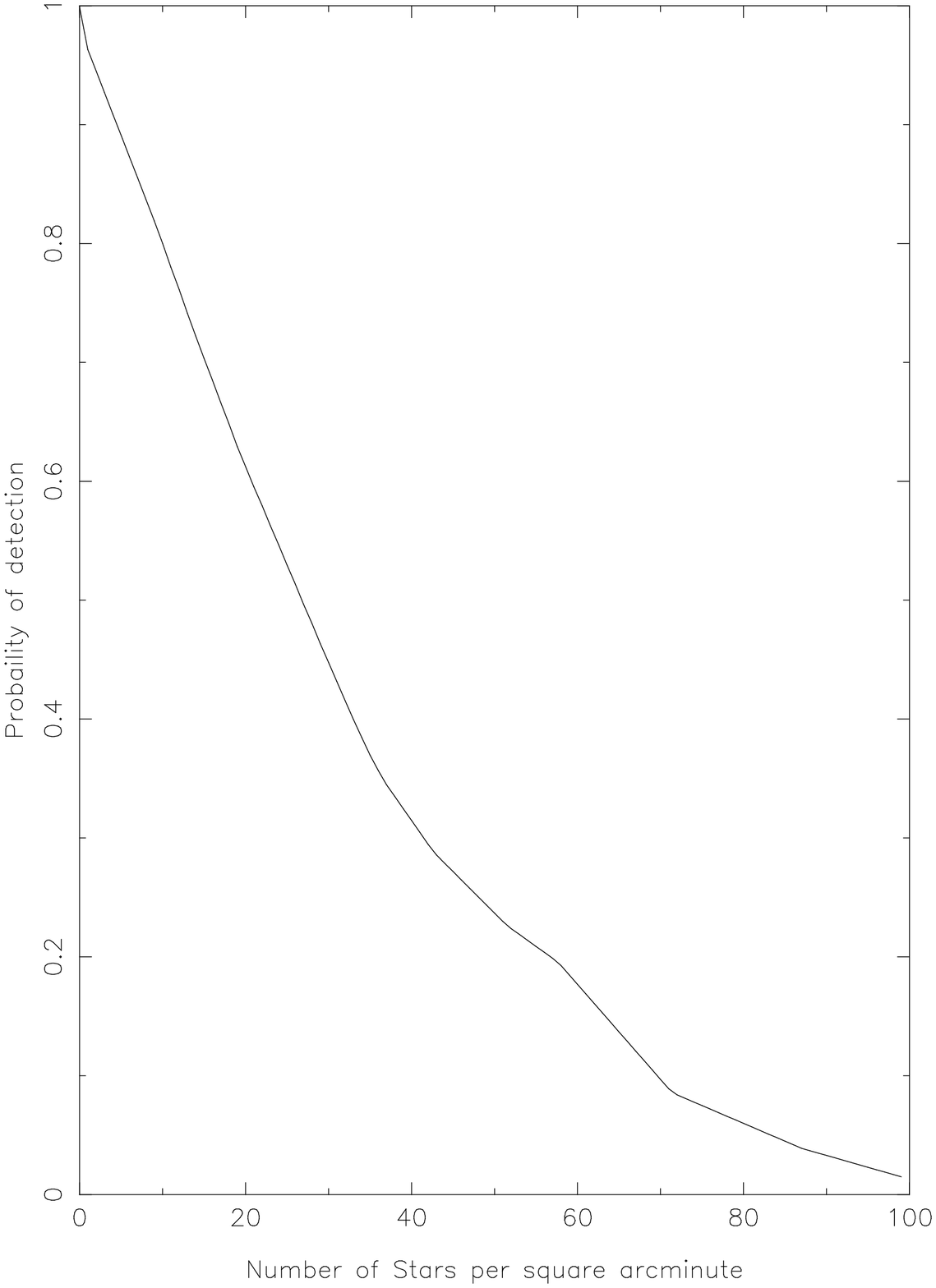}}
      \caption{The probability of detection plotted against the number of 2MASS
images per square arc minute.}
         \label{2MASSprox}
   \end{figure}
\begin{figure*}[htb]     
        \begin{center}
\resizebox{\hsize}{!}{\includegraphics{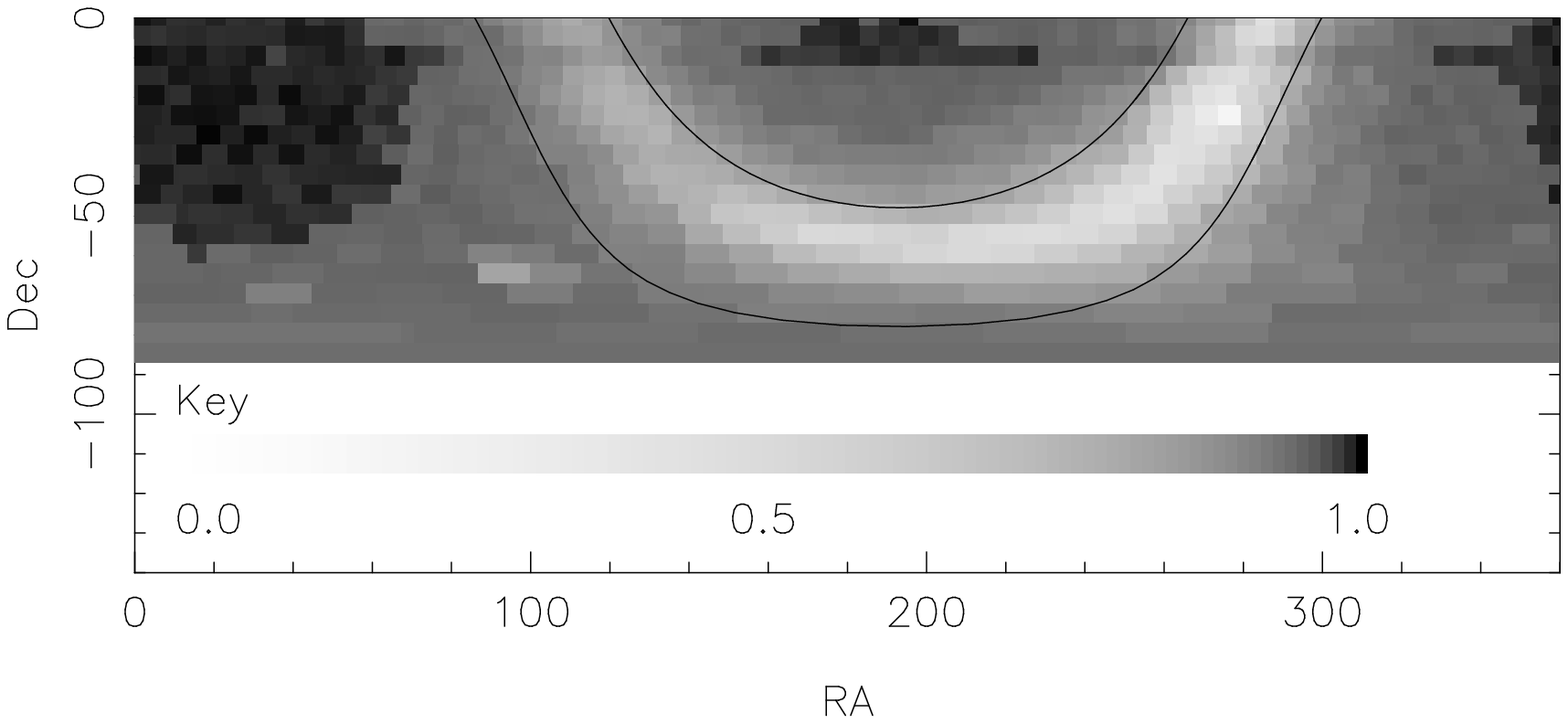}}
\end{center}
\caption{The areal completeness of the survey. The scale shows the
probability that an area is free from both crowding and bright stars. The
solid lines marked represent the galactic latitude cut.}
         \label{areaplot}
\end{figure*}
\subsubsection{SIPS0052$-$6201}
The astrometric solution for this object suggests that it has a common
proper motion with LHS 128. LHS 128 is a K5 dwarf with a parallax of
52 mas (Perryman et al. 1997), if the two objects are part of the same
system this yields a separation perpendicular to the line of sight of
roughly 50,000 AU. It is also possible that the two objects are not
part of a binary system but are simply members of the same moving group.
\section{Completeness}
\subsection{Areal Completeness}                     
As the final aim of this survey is to produce an estimate for the
local space density of late M, L and T Dwarfs we need to analyse the
area covered by our survey. We have already excluded objects near the
Galactic Plane due to crowding. Our other main problem related to
crowding is that of the proximity flag given in the 2MASS Point Source
Catalogue. As advised by the 2MASS Executive Summary we have
excluded all objects within 6 arcseconds of their nearest
neighbour; this will present completeness problems in crowded
regions. In order to produce an estimate as to the magnitude of these
completeness problems we produced a series of simulations which allow
us to calculate the probability of an object having a proximity flag
greater than 6 arcseconds for a given sky surface density of 2MASS
images. To do this a number of images were randomly placed in a one
arcminute square box. If any of these fell with 6 arcseconds of the
centre the test was deemed to have failed. However if no image fell with 6 arcseconds of the
centre the the test was deemed to have succeeded. The probability of
detection is thus the number of successful tests divided by the total
number of tests. Simulations were carried out for different numbers of
stars per square arc minute ranging from 1 to 100. The results of
these simulations are shown in Figure~\ref{2MASSprox}. In order to
calculate the probability of objects being identified in particular
regions of the sky, the number of 2MASS images per square arcminute
and hence the probability of detection were calculated for each UKST
field. \\

In addition there will be areal completeness problems caused by areas
of UKST plates which are near to bright stars. Luckily the size of
these areas is calculated in the post processing of each SuperCOSMOS
scan. Hence we simply multiply this probability of detection
that for the 2MASS images to yield an estimate of the areal
completeness for each field. Figure~\ref{areaplot} shows these data as
a grey-scale plot. It is clear that the galactic plane is crowded and
hence unsuitable for inclusion in this survey.\\

Further completeness issues will need to be addressed before a proper estimate of the the luminosity function (and hence mass function) of the objects included here can be determined. Proper motion completeness will need to be simulated for, with the main factors being the one arcsecond ``movement'' cut and the lower proper motion limit. Also photometric completeness must be estimated taking both the 2MASS colour cut and the limit on the $I$ plate magnitude.  
\subsection{Comparison with other studies}
\begin{figure*}[htb]     
        \begin{center}
\resizebox{\hsize}{!}{\includegraphics{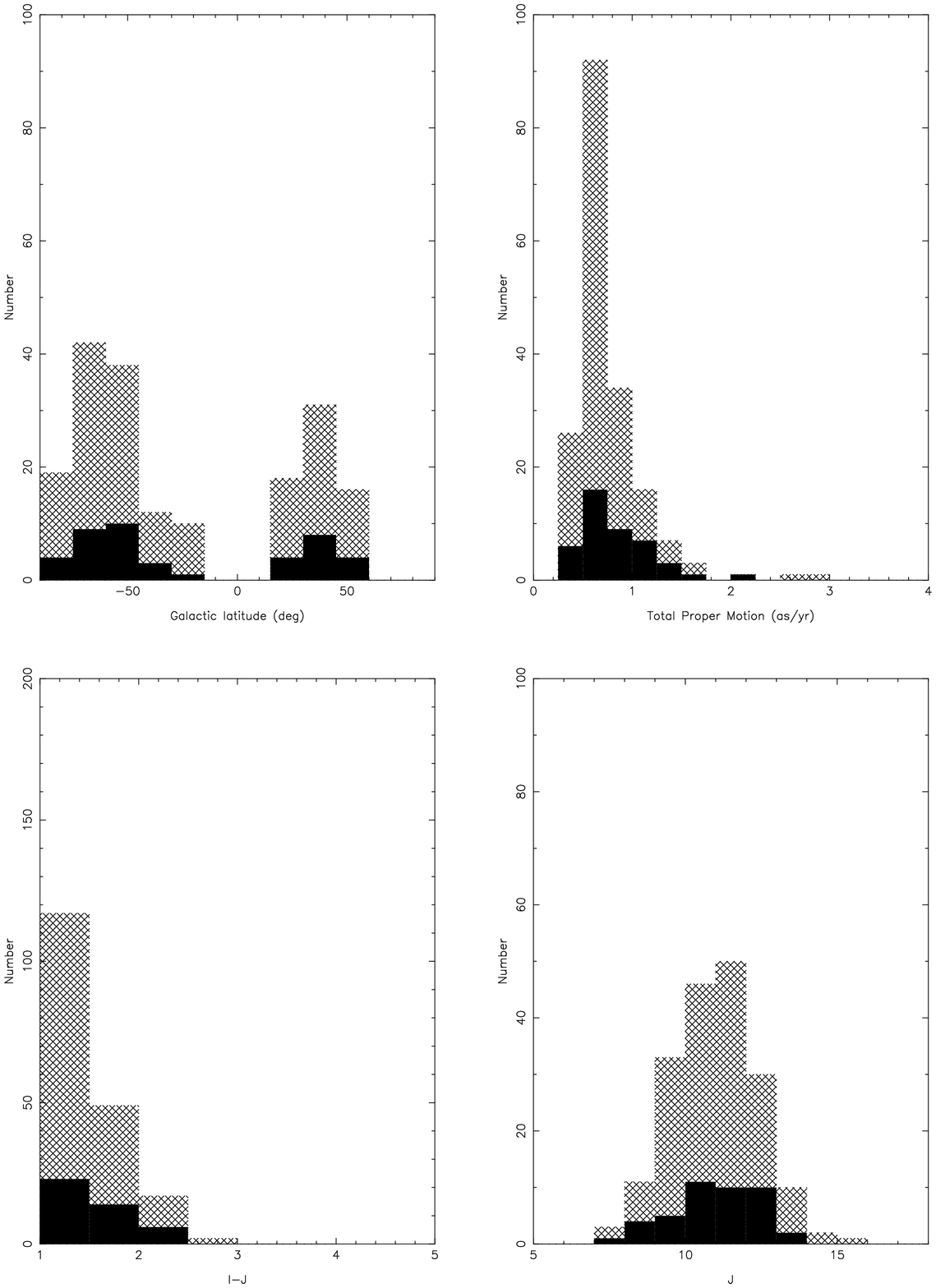}}
\end{center}
\caption{The number of Luyten objects that could possibly be detected (grey)
and those that were (solid).}
         \label{luyten1}
\end{figure*}
In order to make clear the rough extent of the currently unquantified
incompleteness due to astrometric and deblending cuts the survey was
compared to the Luyten Half-Arcsecond Catalogue (Luyten 1979). Using
the accurate positions of Bakos, Sahu and Nemeth (2002) the 2MASS data
for the LHS objects was extracted. These objects were then subjected
to the photometric, crowding and bright star cuts as the survey
data. Figure~\ref{luyten1} shows the number that could be detected and
those that were. It is clear that the survey is somewhat
incomplete, with only 43 of the 200 Luyten objects recovered. This is mainly due to many 2MASS and $I$ observations
being taken at virtually the same time leading to a short proper
motion baseline but deblended and elliptical objects will also play a role. However this is not an accurate estimate of the
incompleteness of the survey. As the Luyten survey will be
unquantifiably incomplete
itself it will be necessary to produce simulations for both the
photometric and astrometric incompleteness in order to gain a complete
picture of the total incompleteness of the survey.
\section{Conclusion and Further Work}
We have utilised wide field surveys to discover 72 new objects with
proper motions greater than 0.5''/yr. Of these objects we have
discovered that one of them has a significant trigonometric parallax
of $\pi = 276 \pm 41$ milliarcseconds yielding a distance of $3.62\pm0.54$pc.\\
In this paper our lower proper motion sets a bias towards nearer and
hence brighter objects. In future papers we hope to detect objects with much lower proper
motions allowing us to expand our search volume greatly. We will also attempt to identify ultracool members of
different kinematic populations by means of reduced proper motions. Once the SuperCOSMOS Sky Survey is
completed on the northern sky we hope to make use of these data
too. Finally we hope to utilise the data yieled to these studies
along with more substantial magnitude, proper motion and volume
completeness estimates to calculate the local space density of late M,L and
T dwarfs.  
\begin{acknowledgements}The authors would like to thank Sue Tritton and Mike Read for their
help in selecting plates, Harvey MacGillivray and Eve Thomson for
their prompt scanning of the material on SuperCOSMOS, David Bacon
for his useful discussions on statistics, Mairi Brookes and Jessica
Skelton for their discussion on spectroscopy and to Stuart Ryder for taking spectroscopic observations. 
This publication makes use of data products from the Two Micron All Sky Survey, which is a joint project of the University of
Massachusetts and the Infrared Processing and Analysis Center/California Institute of Technology, funded by the National
Aeronautics and Space Administration and the National Science Foundation. SuperCOSMOS is funded by a grant from the
UK Particle Physics and Astronomy Research Council.      
\end{acknowledgements}

\clearpage
\newpage
\appendix
\section{Catalogue tables}
\label{Tabs}
Table~\ref{NPos} contains details of objects previously undiscovered
with proper motions greater than 0.5''/yr. Objects with proper motions
taken from the SuperCOSMOS sky survey are marked with ``a'', while
objects with proper motions calculated from 2MASS and SuperCOSMOS $I$
data are marked with ``b''. Objects marked ``c'' lacked sufficient
reference stars to produce a plate to plate fit. For objects with LHS numbers see Luyten
(1979) and references therein. Objects with a CE prefix
can be found in Ruiz et al. (2001), objects with LEHPM
numbers are in Pokorny, Jones \& Hambly (2003) and APMPM
objects are to be found in Scholz et al. (2002). For objects with a quoted 2MASS designation see Burgasser et
al. (2004) and references therein. For objects
with a quoted DENIS designation see Kendall et
al. (2004) and references therein. Finally for objects
marked WT see Wroblewski \& Costa (2001) and references therein.
\begin{table*}[h!!]
   \caption[]{Objects marked * were simultaneously discovered by Subasavage et al. (2005).}
         \label{NPos}
\begin{tabular}{lcccccccc}
\hline
Name & Position (J2000)&Epoch& $\mu_{tot}$ & PA & $\sigma_{\mu}$ & Proper Motion\\
&&&&(arcseconds/yr)&(Degrees)&Source\\
\hline
SIPS0308$-$8212 *& 03 08 53.50 -82 12 34.9 &1990.818& 0.511 &28.9 & 0.017 & b\\ 
SIPS1240$-$8209 *& 12 40 53.21 -82 09 03.9 &1991.100& 0.541 & 276.7 & 0.019 & b\\ 
SIPS2130$-$7710 *& 21 30 06.31 -77 10 36.2 &1995.669& 0.653 & 118.1 & 0.037 & b\\ 
SIPS1633$-$7603 & 16 33 57.58 -76 03 54.4 &1996.397& 0.510 &198.9 & 0.055 & b\\
SIPS2150$-$7520 & 21 50 14.20 -75 20 34.7 &1993.733& 1.019 &106.0&0.050 & b\\ 
SIPS0452$-$7322 *& 04 52 04.61 -73 22 03.6 &1978.930& 0.582 & 54.2 & 0.015 & b\\ 
SIPS0321$-$7046 & 03 21 14.80 -70 46 11.5 &1990.979& 0.553 & 40.3 & 0.025 & b\\ 
SIPS1924$-$6920 & 19 24 35.60 -69 20 13.0 &1996.357& 0.817 & 148.3 & 0.042 & b\\ 
SIPS2032$-$6918 & 20 32 32.62 -69 18 56.4 &1994.650& 0.585 & 147.2 & 0.027 & b\\ 
SIPS1932$-$6506 & 19 32 36.39 -65 06 44.9 &1994.353& 0.510 & 168.3 & 0.030 & b\\ 
SIPS0052$-$6201 & 00 52 14.49 -62 01 55.1 &1994.621&1.123  & 82.8 & 0.050 & b\\ 
SIPS1943$-$6125 & 19 43 33.11 -61 25 37.0 &1993.637& 0.791 & 179.0 & 0.034 & b\\ 
SIPS0523$-$5608 & 05 23 03.82 -56 08 42.5 &1992.033& 0.683 &  341.8& 0.024 & b\\ 
SIPS1936$-$5502 & 19 36 01.69 -55 02 30.7 &1996.559& 0.836 & 133.8 &0.051  & b\\ 
SIPS2053$-$5409 & 20 53 03.74 -54 09 33.7 &1994.688&0.722& 158.1 &0.033 & b\\  
SIPS2242$-$4514 & 22 42 04.80 -45 14 57.1 &2000.731& 0.836  & 280.5 &0.225& b\\ 
SIPS1410$-$4425 & 14 10 41.08 -44 25 55.9 &1982.437&0.537  & 195.9 &0.024 & b\\ 
SIPS1259$-$4336 & 12 59 03.87 -43 36 22.1 &1991.226&1.133&104.0&0.021& b\\ 
SIPS0641$-$4322 & 06 41 18.30 -43 22 36.2 &1994.212&0.689&19.2&0.040& b\\ 
SIPS1337$-$4311 & 13 37 56.25 -43 11 28.2 &1994.453&0.552&227.4&0.046& b\\ 
\hline
\end{tabular}
\end{table*}
\addtocounter{table}{-1}
\begin{table*}
   \caption[]{}
\begin{tabular}{lcccccccc}
\hline
Name & Position (J2000)&Epoch& $\mu_{tot}$ & PA & $\sigma_{\mu}$ & Proper Motion\\
&&&&(arcseconds/yr)&(Degrees)&Source\\
\hline
SIPS1910$-$4132 & 19 10 33.49 -41 32 38.8 &1983.346&0.738 & 174.7 & 0.016 & b\\ 
SIPS1231$-$4018 & 12 31 21.55 -40 18 37.1 &1993.249&0.679&274.0&0.033& b\\ 
SIPS1149$-$4012 & 11 49 50.12 -40 12 42.0 &1991.223&0.566&253.2&0.034& b\\ 
SIPS0013$-$3921 & 00 13 00.54 -39 21 19.9 &1994.648&0.522&102.7&0.033& b\\ 
SIPS1251$-$3846 & 12 51 32.08 -38 46 13.1 &1994.297&0.593&278.9&0.036& b\\ 
SIPS0031$-$3840 & 00 31 18.84 -38 40 35.1 &1990.817&0.568&96.8&0.035& c\\ 
SIPS1039$-$3819 & 10 39 32.57 -38 19 57.2 &1992.080&0.724&203.6&0.035& b\\ 
SIPS1338$-$3752 & 13 38 28.14 -37 52 50.2 &1984.304&1.278&269.6&0.014& b\\ 
SIPS1141$-$3624 & 11 41 21.05 -36 24 38.3 &1988.154&0.556&57.1&0.013& b\\ 
SIPS1541$-$3609 & 15 41 19.25 -36 09 10.5 &1994.407&0.507&263.7&0.038& b\\ 
SIPS1342$-$3534 & 13 42 22.47 -35 34 47.9 &1984.304&0.931&259.1&0.015& b\\  
SIPS2050$-$3358 & 20 50 44.01 -33 58 37.5 &1986.563&0.510&120.4&0.022& b\\ 
SIPS0417$-$3211 & 04 17 58.82 -32 11 49.4 &2000.967&0.855&283.7&0.111& b\\ 
SIPS2346$-$3153 & 23 46 54.50 -31 53 50.5 &1992.551&0.649&134.9&0.043& b\\ 
SIPS0405$-$3138 & 04 05 29.44 -31 38 44.6 &2000.967&0.846&261.2&0.053& b\\ 
SIPS2314$-$2929 & 23 14 26.40 -29 29 52.8 &1994.569& 0.620 & 132.7 & 0.043 & a\\
SIPS1529$-$2907 & 15 29 14.08 -29 07 28.8 &1991.452&1.078 &185.8&0.025& b\\ 
SIPS1548$-$2859 & 15 48 21.55 -28 59 33.9 &1994.349&0.546 &219.0&0.052& b\\ 
SIPS2313$-$2826 & 23 13 59.28 -28 26 36.9 &1994.653&0.825 &117.5 &0.054 & b\\ 
SIPS0422$-$2802 & 04 22 31.25 -28 02 30.6 &2000.967& 0.914&281.5&0.077& b\\ 
SIPS2231$-$2756 & 22 31 22.38 -27 56 47.3 &1999.691& 1.677 & 105.8 & 0.407 & b\\ 
SIPS1025$-$2730 & 10 25 55.01 -27 30 59.6 &1996.227& 0.502 & 223.9 & 0.041& b\\ 
SIPS2157$-$2726 & 21 57 54.53 -27 26 28.8 &1994.692& 0.855 & 131.4 &0.053  & b\\ 
SIPS2308$-$2721 & 23 08 11.23 -27 21 59.1 &1994.771& 0.597 &124.7  & 0.067 & b\\ 
SIPS1019$-$2707 & 10 19 24.68 -27 07 17.3 &1994.371& 0.582  &272.3 &0.044 & b\\ 
SIPS1540$-$2613 & 15 40 30.18 -26 13 35.4 &1993.328& 1.623 & 224.9 & 0.021& b\\ 
SIPS1424$-$2535 & 14 24 53.94 -25 35 17.9 &1997.238& 0.641 &105.5 &0.175& b\\ 
SIPS1627$-$2518 & 16 27 49.75 -25 18 13.5 &1982.568& 0.582 & 210.8 & 0.015 & b\\ 
SIPS1513$-$2243 & 15 13 59.26 -22 43 47.6 &1992.242& 0.574 & 206.4 & 0.026 & b\\ 
SIPS1250$-$2121 & 12 50 52.43 -21 21 11.4 &1993.399& 0.578 & 126.0 &0.042& c\\ 
SIPS0921$-$2104 & 09 21 13.97 -21 04 38.1 &1993.169& 0.965  &162.6 &0.016 & b\\ 
SIPS0004$-$2058 & 00 04 41.17 -20 58 30.3 &1994.733& 0.826 & 90.6& 0.107 & a\\ 
SIPS2359$-$2007 & 23 59 57.48 -20 07 37.6 &1994.733& 0.798 &127.7 &0.040& b\\ 
SIPS2210$-$1952 & 22 10 50.04 -19 52 22.1 &1994.771& 0.830 &188.6 &0.051 & b\\ 
SIPS1048$-$1925 & 10 48 18.98 -19 25 33.3 &1996.388& 0.957 &264.4 &0.112& b\\ 
SIPS1121$-$1653 & 11 21 09.57 -16 53 52.1 &1993.396& 0.602 &204.8 &0.0473 & b\\ 
SIPS0227$-$1624 & 02 27 10.17 -16 24 46.2 &1994.927& 0.592 & 120.8 &0.019  & b\\ 
SIPS1109$-$1606 & 11 09 27.58 -16 06 50.4 &1994.139& 0.520 & 233.9 & 0.050 & b\\ 
SIPS0933$-$1602 & 09 33 49.78 -16 02 50.7 &1997.000& 1.357& 188.3 & 0.072& b\\ 
SIPS0050$-$1538 & 00 50 24.35 -15 38 20.0 &2001.796& 0.674 & 227.3 & 0.042 & b\\ 
SIPS0258$-$1220 & 02 58 35.03 -12 20 18.3 &2001.805& 0.827 & 91.4 & 0.042 & b\\ 
SIPS1413$-$1201 & 14 13 05.33 -12 01 22.4 &1988.545& 0.715&237.2&0.025& b\\ 
SIPS1542$-$1007 & 15 42 25.44 -10 07 02.3 &1996.302& 0.552&166.0&0.062&b\\ 
SIPS2025$-$0835 & 20 25 47.58 -08 35 29.0 &1994.661& 0.578&151.8&0.052& b\\ 
SIPS0320$-$0446 & 03 20 28.43 -04 46 33.4 &1994.698& 0.678 &190.6 &0.038& b\\
SIPS1956$-$0422 & 19 56 50.17 -04 22 39.3 &1993.634& 0.563&115.0&0.026& b\\ 
SIPS0820$-$0355 & 08 20 44.43 -03 55 04.5 &1996.224& 0.594&146.6&0.123& b\\ 
SIPS0251$-$0352 & 02 51 14.53 -03 52 36.3 &1993.690& 2.185&149.3&0.057& b\\ 
SIPS0346$-$0218 & 03 46 11.89 -02 18 18.1 &1986.891& 0.850&155.4&0.021& b\\ 
SIPS0937$-$0214 & 09 37 31.35 -02 14 32.4 &1987.099& 0.528&127.6&0.025& b\\ 
SIPS2343$-$0155 & 23 43 04.40 -01 55 06.0 &1999.786& 1.359&279.4&0.288& b\\ 
SIPS1512$-$0112 & 15 12 17.92 -01 12 28.3 &1990.489& 0.671 &191.6&0.031& b\\
\hline
\end{tabular}
\end{table*}
\begin{table*}
   \caption[]{}
         \label{PPos}
\begin{tabular}{lcccccccc}
\hline
Name & Position (J2000)&Epoch& $\mu_{tot}$ & PA & $\sigma_{\mu}$ & Proper Motion\\
&&&&(arcseconds/yr)&(Degrees)&Source\\
\hline 
SCR 1845$-$6357& 18 45 02.32 -63 57 52.6 &1992.586 & 2.541 & 74.1 & 0.153 & a\\ 
LHS 532& 22 56 25.04 -60 03 44.4 &1994.648&1.020&206.6&0.024 & b\\ 
LHS 3492 & 19 51 31.35 -50 55 33.1 &1994.366& 0.815 & 192.2 &0.033 &b\\
APMPM J04052-5819& 04 51 37.56 -58 18 44.5 &1989.038& 0.701 & 192.6 & 0.025 & b\\ 
WT 60& 01 52 00.00 -57 47 55.7 &1994.858& 0.636 & 216.5 & 0.037 & b\\ 
$\epsilon$ Indi B& 22 04 09.46 -56 46 52.6 &1997.771& 4.840 & 120.8 &0.048& c\\ 
APMPM J0524$-$5607& 05 23 30.59 -56 07 06.4 &1992.033& 0.782 & 5.8 & 0.029& b\\ 
LEHPM 1683& 01 34 59.74 -55 11 34.9 &1994.856& 0.644 & 135.0 & 0.042 & b\\ 
APMPM J 0247$-$5257& 02 47 20.63 -52 56 45.9 &1997.723& 0.776 & 47.1 & 0.102 & a\\ 
LHS 1471& 02 55 14.27 -51 40 21.8 &1996.964& 0.656 & 70.5 & 0.094 & a\\ 
LEHPM 3396& 03 34 08.52 -49 53 39.5 &1984.915& 2.448 & 77.6 & 0.055 & a\\ 
DENIS J0255$-$4700& 02 55 03.18 -47 00 48.6 &1994.809& 1.197 & 120.5 & 0.126 & a\\ 
LHS 3971& 23 34 15.53 -40 44 03.8 &1984.625& 0.857& 105.2 & 0.023 & a\\ 
LHS 1502& 03 08 49.00 -40 06 40.4 &1995.784&0.739&142.9&0.058& b\\ 
DENIS J1048$-$3956& 10 48 15.39 -39 55 59.3 &1992.080&1.541&231.0&0.013& b\\ 
LHS 1490& 03 02 06.56 -39 50 49.3 &1995.784&0.792&221.5&0.025& b\\ 
LHS 539& 23 15 51.15 -37 33 32.1 &1995.625& 1.352 &  71.1&0.084& b\\ 
2MASSW J1155395$-$372735& 11 55 39.51 -37 27 32.5 &1996.189&0.868&172.5&0.039& b\\ 
LHS 1531& 03 19 17.03 -37 03 42.0 &1995.989&0.818&124.8&0.039&b\\ 
LEHPM 3070& 03 06 11.65 -36 47 50.3 &1995.989& 0.571 & 179.8 & 0.243 & a\\ 
CE 136& 10 11 13.49 -35 36 21.6 &1993.325&0.509&293.2&0.030& b\\ 
LHS 1192& 01 05 52.14 -34 34 48.5 &1991.582&0.607&92.2&0.035& b\\ 
LHS 1355& 02 12 39.83 -33 52 06.1 &2000.723&1.045 &93.5  &0.083 & b\\  
LHS 2506& 12 03 58.50 -33 01 26.8 &1996.189&0.768&275.4&0.033& b\\ 
CE 220& 10 45 34.15 -32 49 53.5 &1993.246&0.551&306.6&0.046& b\\ 
APMPM J1407$-$3018& 14 06 49.79 -30 18 27.6 &1993.208&0.852&264.3&0.028& b\\ 
LHS 3566& 20 39 23.68 -29 26 30.6 &1994.538& 0.832 & 154.1 &0.041 & b\\ 
LHS 3662& 21 23 07.74 -28 09 54.4 &1994.773& 0.715 &132.9&0.057 & b\\ 
APMPM J1224$-$2758& 12 23 56.82 -27 57 48.4 &1993.243&1.247 &282.9 &0.027 & b\\ 
APMPM J2331$-$2750& 23 31 21.70 -27 49 53.6 &1994.569& 0.781 & 1.439 & 0.136 & a\\ 
LHS 2487& 11 58 57.02 -27 43 07.5 &1993.246&0.501&295.4&0.031& b\\ 
DENIS J1456$-$2747& 14 56 01.44 -27 47 34.7 &1997.361&0.821  & 197.3& 0.028& b\\ 
LHS 2597& 12 39 36.56 -26 58 10.6 &1997.361& 0.713 &259.5  &0.036  & b\\ 
LHS 2928& 14 30 18.06 -24 03 15.3 &1997.238& 0.637 & 228.8 & 0.062 & a\\   
LEHPM 162& 00 05 48.18 -21 57 19.0 &1994.733& 0.778 & 104.7&0.045 & b\\ 
LHS 3970& 23 33 40.41 -21 33 51.9 &1995.628& 0.794 &103.2 &0.027 & b\\ 
LHS 1069& 00 24 39.49 -20 54 09.1 &1995.669& 0.507 &198.7& 0.042 & b\\ 
LHS 105& 00 09 16.95 -19 42 34.5 &1994.733&1.093 &60.8 &0.036 & b\\ 
LHS 3718& 21 49 23.86 -18 10 36.6 &1994.618&0.590  &206.4 &0.052  & b\\ 
LHS 3799& 22 23 06.98 -17 36 26.0 &1999.606&0.929&164.5 &0.240& c\\ 
LHS 138& 01 12 30.28 -16 59 59.1 &1995.822& 1.386 & 63.6 & 0.160 & a\\  
2MASS J15074769$-$1627386& 15 07 47.70 -16 27 37.5 &1997.244& 1.038 & 182.4 &0.255& c\\ 
LHS 1396& 02 23 01.78 -16 16 43.5 &1994.927&0.781 &125.7 &0.033 & b\\ 
LHS 2397& 11 20 26.46 -14 39 59.8 &1993.396&0.509 &215.1&0.038& b\\ 
LHS 1462& 02 52 39.40 -14 31 44.6 &2001.805& 0.865 & 195.7 & 0.042 & a\\ 
LHS 500& 20 55 35.92 -14 03 49.2 &1987.734&1.484  & 107.8 &0.013  & b\\ 
LHS 523& 22 28 54.36 -13 25 19.1 &1999.601& 1.222 &209.1 &0.129 & b\\ 
LHS 3056& 15 19 11.90 -12 45 05.6 &1996.394&0.772 &236.6 & 0.077& b\\ 
LHS 3935& 23 21 13.24 -11 37 23.1 &1994.776&0.760&104.3&0.048& b\\ 
LHS 3911& 23 13 19.35 -11 06 17.1 &1994.776&0.926&101.9&0.072& b\\ 
2MASSW J1555157$-$095605& 15 55 15.28 -09 55 59.8 &1992.310&1.273&131.1&0.022& b\\ 
\hline
\end{tabular}
\end{table*}
\addtocounter{table}{-1}
\begin{table*}
   \caption[]{}
\begin{tabular}{lcccccccc}
\hline
Name & Position (J2000)&Epoch& $\mu_{tot}$ & PA & $\sigma_{\mu}$ & Proper Motion\\
&&&&(arcseconds/yr)&(Degrees)&Source\\
\hline
LHS 3735& 21 57 39.27 -09 28 13.6 &2000.723&0.541&248.9&0.061& b\\ 
LHS 353& 13 30 03.00 -08 42 24.3 &1997.320&1.362&249.2&0.041& b\\ 
LHS 176& 03 35 38.61 -08 29 22.9 &1999.995&1.748&112.2&0.050& b\\ 
LHS 424& 16 55 35.58 -08 23 35.1 &1993.632&1.163&221.503&0.031& b\\ 
LHS 2& 00 06 43.16 -07 32 18.5 &2000.723&2.063&201.3&0.120& c\\ 
LHS 5217& 12 31 23.99 -06 38 00.2 &1997.238&0.533&117.0&0.045& b\\ 
LHS 3149& 16 04 20.02 -06 16 39.6 &1992.504&0.867&182.9&0.028& b\\ 
DENIS J1539$-$0520& 15 39 41.61 -05 20 43.3 &1992.506&0.644&83.3&0.042& c\\ 
LHS 3611& 21 01 53.75 -05 23 40.1 &1995.510& 0.528 & 179.6 & 0.025 & a\\ 
LHS 1785& 05 47 08.98 -05 12 09.3 &1996.068& 0.787 & 136.4 & 0.037 & a\\ 
LHS 2419& 11 31 26.85 -05 03 36.3 &1995.138& 0.582 & 278.7 & 0.029 & a\\ 
2MASS J23062928$-$0502285& 23 06 28.98 -05 02 26.0 &1993.730&1.042&119.7&0.025& b\\ 
LHS 17& 02 46 14.21 -04 59 08.8 &1993.690&2.488&137.9&0.041& b\\ 
DENIS J0012$-$0457& 00 13 46.38 -04 57 36.2 &1994.651&0.685&107.4&0.039& b\\ 
LHS 1363& 02 14 12.38 -03 57 42.8 &1994.659&0.504&105.0&0.079& b\\ 
LHS 2641& 12 49 48.72 -03 17 32.3 &1997.243&0.608&195.6&0.049& a\\ 
LHS 3104& 15 42 30.00 -03 15 51.7 &1992.506&0.504&253.7&0.027& b\\ 
LHS 1332& 02 04 27.97 -01 52 49.9 &1987.822&0.799&225.2&0.013& b\\ 
LHS 1415& 02 50 06.29 -01 50 32.8 &1986.932&0.801&76.9&0.021& b\\ 
LHS 5165& 10 03 19.26 -01 05 08.1 &1997.123&0.538&284.0&0.060& b\\ 
LSR J1610$-$0040& 16 10 28.93 -00 40 54.3 &2000.242&1.719&226.6&0.088& b\\
\hline
\end{tabular}
\end{table*}
   \begin{table*}[h!!]
      \caption[]{}
         \label{NPhot}
\begin{center}
\begin{tabular}{ccccc}
\hline
Object & $I$ & $J$ & $H$ & $K_{S}$\\
\hline
SIPS0308$-$8212 &13.087&11.704&11.145&10.894\\  
SIPS1240$-$8209 &12.490&10.855&10.196&9.933\\ 
SIPS2130$-$7710 &13.476&11.292&10.666&10.365 \\
SIPS1633$-$7603 &16.655&14.261&13.644&13.298 \\
SIPS2150$-$7520 &17.260&14.056&13.176&12.673 \\
SIPS0452$-$7322 &13.621&11.976&11.442&11.124 \\
SIPS0321$-$7046 &11.746&10.394&9.840&9.574\\
SIPS1924$-$6920 &14.267&12.392&11.908&11.607 \\
SIPS2032$-$6918 &16.298&13.636&12.983&12.579 \\
SIPS1932$-$6506 &14.784&12.664&12.048&11.719 \\
SIPS0052$-$6201 &14.184&12.153&11.742&11.370 \\
SIPS1943$-$6125 &16.115&13.199&12.700&12.287 \\
SIPS0523$-$5608 &15.580&13.554&12.948&12.602 \\
SIPS1936$-$5502 &17.474&14.486&13.628&13.046 \\
SIPS2053$-$5409 &13.916&12.170&11.656&11.345 \\
SIPS2242$-$4514 &17.501&15.981&15.586&15.120 \\
SIPS1410$-$4425 &17.980&15.874&15.066&14.544 \\
SIPS1259$-$4336 &12.799&10.534&9.954&9.520 \\
SIPS0641$-$4322 &17.170&13.751&12.941&12.451 \\
SIPS1337$-$4311 &12.850&11.701&11.112&10.794 \\
SIPS1910$-$4132 &12.577&11.147&10.552&10.249 \\
SIPS1231$-$4018 &11.538&10.444&9.940&9.635 \\
SIPS1149$-$4012 &17.269&13.999&13.322&12.834 \\
SIPS0013$-$3921 &17.551&14.787&14.216&13.809 \\
SIPS1251$-$3846 &13.758&12.086&11.558&11.246 \\
SIPS0031$-$3840 &17.442&14.101&13.399&12.924 \\ 
SIPS1039$-$3819 &12.535&10.846&10.301&9.998 \\ 
SIPS1338$-$3752 &13.141&11.747&11.281&10.965 \\
SIPS1141$-$3624 &10.108&8.490&7.967&7.699 \\ 
SIPS1541$-$3609 &14.467&11.970&11.450&11.109 \\
SIPS1342$-$3534 &14.790&13.755&13.216&12.935 \\
SIPS2050$-$3358 &15.858&13.815&13.146&12.854 \\
SIPS0417$-$3211 &17.562&14.864&14.196&13.739 \\
SIPS2346$-$3153 &15.838&13.279&12.680&12.198 \\
SIPS0405$-$3138 &13.978&12.924&12.282&11.989 \\
SIPS2314$-$2929 &17.329&14.608&14.101&13.682 \\
SIPS1529$-$2907 &15.475&13.319&12.855&12.499 \\
SIPS1548$-$2859 &15.468&13.118&12.551&12.260 \\
SIPS2313$-$2826 &14.499&12.660&12.149&11.878 \\
SIPS0422$-$2802 &16.250&14.755&14.091&13.775 \\
SIPS2231$-$2756 &17.859&15.808&15.318&14.903 \\
SIPS1025$-$2730 &16.699&14.551&14.048&13.625 \\
SIPS2157$-$2726 &17.914&14.672&13.887&13.401 \\
SIPS2308$-$2721 &17.819&14.662&13.833&13.332 \\
SIPS1019$-$2707 &16.753&13.526&12.906&12.471 \\
SIPS1540$-$2613 &14.600&11.646&11.145&10.730\\
SIPS1424$-$2535 &17.751&15.665&15.123&14.784 \\
SIPS1627$-$2518 &17.706&14.888&14.368&14.016 \\
SIPS1513$-$2243 &17.064&14.196&13.601&13.231 \\
SIPS1250$-$2121 &13.336&11.160&10.550&10.128 \\
SIPS0921$-$2104 &16.031&12.779&12.152&11.690 \\
SIPS0004$-$2058 &15.647&12.404&11.834&11.396 \\
SIPS2359$-$2007 &17.380&14.382&13.623&13.248 \\
SIPS2210$-$1952 &16.456&14.000&13.498&13.151\\
SIPS1048$-$1925 &17.830&14.876&14.276&13.692 \\
SIPS1121$-$1653 &13.856&12.553&12.073&11.762 \\
SIPS0227$-$1624 &16.711&13.573&12.630&12.143 \\
\hline
\end{tabular}
\end{center}
   \end{table*}
   \begin{table*}[h!!]
      \caption[]{}
\begin{center}
\begin{tabular}{ccccc}
\hline
Object & $I$ & $J$ & $H$ & $K_{S}$\\
\hline
SIPS1109$-$1606 &17.856&14.970&14.348&13.892 \\
SIPS0933$-$1602 &15.307&12.720&12.232&11.890 \\
SIPS0050$-$1538 &16.880&13.779&13.077&12.647 \\
SIPS0258$-$1220 &15.822&13.921&13.410&13.030 \\
SIPS1413$-$1201 &10.446&9.040&8.453&8.163 \\
SIPS1542$-$1007 &16.859&14.372&13.801&13.437 \\
SIPS2025$-$0835 &15.699&13.153&12.403&12.029 \\
SIPS0320$-$0446 &16.558&13.259&12.535&12.134 \\
SIPS1956$-$0422 &16.949&13.441&12.921&12.510 \\
SIPS0820$-$0355 &17.319&15.884&15.257&15.025 \\ 
SIPS0251$-$0352 &16.214&13.059&12.254&11.662 \\
SIPS0346$-$0218 &15.255&13.130&12.642&12.346 \\
SIPS0937$-$0214 &16.979&14.410&13.883&13.585 \\
SIPS2343$-$0155 &17.868&15.598&15.044&14.757 \\
SIPS1512$-$0112 &16.607&15.244&14.873&14.469 \\
\hline
\end{tabular}
\end{center}
   \end{table*}
   \begin{table*}[h!!]
      \caption[]{}
         \label{PPhot}
\begin{center}
\begin{tabular}{ccccc}
\hline
Object & $I$ & $J$ & $H$ & $K_{S}$\\
\hline
SCR1845$-$6357&12.528&9.544&8.967&8.508\\
LHS 532&10.653&8.984&8.360&8.108\\
LHS 3492 &12.288&10.970&10.493&10.188 \\
APMPM J0452$-$5819&13.980&11.691&11.089&10.705\\
WT 60&13.213&11.450&10.846&10.568\\
$\epsilon$ Indi B&16.567&11.908&11.306&11.208\\
APMPM J0524$-$5607 &13.960&12.337&11.715&11.370\\
LEHPM 1683&15.211&13.227&12.709&12.427\\
APMPM J0247$-$5257&13.097&11.748&11.184&10.873\\
LHS 1471&11.125&9.938&9.366&9.082\\
LEHPM 3396&13.916&11.376&10.823&10.392\\
DENIS 0255$-$4700&17.025&13.246&12.204&11.558\\
LHS 3971&12.683&11.248&10.713&10.435\\
LHS 1502&13.925&12.623&12.105&11.840\\
DENIS 1048$-$3956&12.097&9.538&8.905&8.447\\
LHS 1490&11.926&10.705&10.178&9.885\\
LHS 539&11.556&10.403&9.872&9.592\\
2MASSW J1155395$-$372735&16.370&12.811&12.040&11.462\\
LHS 1531&13.367&12.160&11.676&11.344\\
LEHPM 3070&13.764&11.690&11.068&10.631\\
CE 136&12.750&11.666&11.070&10.807\\
LHS 1192&12.808&11.160&10.569&10.272\\
LHS 1355&12.668&11.141&10.625&10.319\\
LHS 2506&10.618&9.529&9.073&8.744\\
CE 220&11.055&9.9869&9.233&8.966\\
APMPM J1407$-$3018&13.538&11.361&10.687&10.366\\
LHS 3566&13.682&11.357&10.743&10.367\\
LHS 3662&12.589&11.062&10.523&10.247\\
APMPM J1224$-$2758&14.151&11.977&11.401&11.069\\
APMPM J2331$-$2750&14.404&11.646&11.055&10.651\\
LHS 2487&14.359&12.763&12.238&11.963\\
DENIS 1456$-$2747&16.503&13.250&12.655&12.189\\
LHS 2597&11.424&10.042&9.581&9.233\\
LHS 2928&11.857&10.742&10.237&9.917\\
LEHPM 162&16.461&13.274&12.617&12.201\\
LHS 3970&13.725&11.858&11.312&10.929\\
LHS 1069&14.627&12.873&12.312&12.014\\
LHS 105&12.532&10.883&10.327&10.074\\
LHS 3718&13.994&12.532&12.052&11.747\\
LHS 3799&9.268&8.242&7.638&7.319\\
LHS 138&8.523&7.258&6.749&6.420\\
2MASS J15074769$-$1627386&16.134&12.830&11.895&11.312\\
LHS 1396&15.159&13.490&13.027&12.645\\
LHS 2397&14.877&13.444&12.903&12.641\\
LHS 1462&14.094&12.695&12.227&11.900\\
LHS 500&11.248&9.717&9.218&8.915\\
LHS 523&12.639&10.768&10.217&9.843\\
LHS 3056&9.640&8.507&7.862&7.582\\
LHS 3935&13.986&12.296&11.761&11.512\\
LHS 3911&12.866&11.125&10.607&10.299\\
2MASSW J1555157$-$095605&16.057&12.557&11.984&11.443\\
\hline
\end{tabular}
\end{center}
   \end{table*}
   \begin{table*}[h!!]
      \caption[]{}
\begin{center}
\begin{tabular}{ccccc}
\hline
Object & $I$ & $J$ & $H$ & $K_{S}$\\
\hline
LHS 3735&12.019&10.450&9.822&9.537\\
LHS 353&10.731&9.599&9.047&8.749\\
LHS 176&11.596&10.377&9.801&9.456\\
LHS 424&11.595&9.776&9.201&8.816\\
LHS 2&9.876&8.323&7.792&7.439\\
LHS 5217&12.527&11.328&10.747&10.434\\
LHS 3149&11.683&10.452&9.880&9.548\\
DENIS 1539$-$0520&17.805&13.922&13.060&12.575\\
LHS 3611&14.00&12.839&12.33&12.006\\
LHS 1785&11.096&10.039&9.514&9.177\\
LHS 2419&15.216&13.682&13.171&12.860\\
2MASS J23062928$-$0502285&13.580&11.354&10.718&10.296\\
DENIS 0012$-$0457&13.925&11.462&10.866&10.479\\
LHS 17&12.102&10.970&10.499&10.152\\
LHS 1363&11.956&10.481&9.858&9.485\\
LHS 2641&11.874&10.759&10.261&9.925\\
LHS 3104&14.941&12.657&12.162&11.810\\
LHS 1332&10.744&9.585&9.092&8.804\\
LHS 1415&14.520&12.914&12.430&12.119\\
LHS 5165&14.774&12.327&11.667&11.236\\
LSR J1610-0040&14.903&12.911&12.302&12.019\\
\hline
\end{tabular}
\end{center}
   \end{table*}

\begin{thebibliography}{}
\bibitem{bakos} Bakos, G., Sahu, K.C., Nemeth, P., AJSS, 141, 1, 187, 2002
\bibitem{Burgasser2004} 
Burgasser, A.J., McElwain, M.W., Kirkpatrick, J.D., Cruz,
K.L., Tinney, C.G., Reid, I.N., AJ, 127, 5, 2856, 2004
\bibitem{Burgasser2002} Burgasser, A.J. et al., ApJ, 564, , 421, 2002
\bibitem{Dahn} Dahn, C.C. et al., AJ, 124, 2, 1170, 2002 
\bibitem{Deacon2001} Deacon, N.R, Hambly, N.C., A\&A, 380, 148, 2001
\bibitem{Deacon2004} Deacon, N.R, Hambly, N.C., Henry, T.J.,
Subasavage, J., Brown, M., Jao, W.C., AJ accepted, astro-ph/0409582, 2005
\bibitem{Geballe} Geballe, T.R. et al. ApJ, 564, 1, 466, 2002
\bibitem{Gorlova} Gorlova, N. I., Meyer, M. R., Rieke, G. H., Liebert,
J., ApJ, 593, 2, 1074, 2003
\bibitem{Hambly2001a}Hambly, N.C., MacGillivray, H.T., Read, M.A.,
Tritton, S.B., Thomson, E.B., Kelly, B.D., Morgan, D.H., Smith, R.E.,
Driver, S.P., Williamson, J., Parker, Q.A., Hawkins, M.R.S., Williams,
P.M., Lawrence, A.,MNRAS, 326, 4, 1279, 2001
\bibitem{Hambly2001c}Hambly, N. C.; Davenhall, A. C.; Irwin, M. J.;
MacGillivray, H. T., MNRAS, 326, 4, 1315, 2001
\bibitem{Hambly2003} Hambly, N.C.,
Henry, T.J., Subasavage, J., Brown, M., Jao, W.C., AJ, 128:437-447, 2004
\bibitem{Kendall}
Kendall, T. R., Delfosse, X., Martin, E. L., Forveille, T., A\&A 416,
L17, 2004
\bibitem{Kirkpatrick} Kirkpatrick, J.D., Reid, I.N,, Liebert, J.,
Gizis, J.E., Burgasser, A.J., Monet, D.G., Dahn, C., Nelson, B.,
Williams, R.J.,AJ, 120, 1, 447, 2000
\bibitem{Lepine} Lepine, S., Shara,M.M., AJ, 124, 2002
\bibitem{Luyten} Luyten Half Arcsecond Catalogue, Luyten, W.J.,
University of Minnesota, Minneaplois, 1979
\bibitem{Perryman} Perryman, M. A. C., Lindegren, L., Kovalevsky, J.,
Hoeg, E., Bastian, U., Bernacca, P. L., Crézé, M., Donati, F., Grenon,
M., van Leeuwen, F., van der Marel, H., Mignard, F., Murray, C. A., Le
Poole, R. S., Schrijver, H., Turon, C., Arenou, F., Froeschlé, M.,
Petersen, C. S.,  A\&A, 323, L49, 1997
\bibitem{Pokorny} Pokorny, R.S., Jones, H.R.A., Hambly, N.C., A\&A, 397,
575-584, 2003
\bibitem{Ruiz} Ruiz, M.T., Wischnjewsky, M., Rojo, P.M., Gonzalez,
L.E., ApJSS, 133, 119-160, 2001
\bibitem{Subasavage} Subasavage, J.P., Henry, T.J., Hambly, N.C.,
Brown, M.A., Jao, W.C., AJ submitted, 2005
\bibitem{Scholz 2002} Scholz, R.D., Ibata, R., Irwin, M., Lehmann, I., Salvato, M.,
Schweitzer, A., MNRAS, 329, 109-114, 2002
\bibitem{Scholz 2003a} Scholz, R.D., McCaughrean, M.J., Lodieu, N., Kuhlbrodt, B., A\&A, v. 398,
L29-L33, 2004 
\bibitem{Seifahrt} Seifahrt, A., Neuhäuser, R., Mugrauer, M., A \& A, 421, 255, 2004
\bibitem{Teegarden} Teegarden, B.J., Pravdo, S.H., Hicks, M., Lawrence, K., Shaklan,
S.B., Covey, K., Fraser, O., Hawley, S.L., McGlynn, T., Reid, I.N., ApJ, 589,
L51-L53, 2003

\bibitem{Vrba} Vbra, F.J. et al., AJ, 127, 5, 2948, 2004

   \bibitem[1998]{Wallace} Wallace, P.T., Starlink User Note
   No. 67.42: SLALIB: Postitional Astronomy Library, CCLRC/Rutherford Appleton
   Laboratory, PPARC, 1998
\bibitem{Wroblewski} Wroblewski, H.,
Costa, E., A\&A, 367, 725-728, 2001

\bibitem{2MASSsum} The 2MASS Point Source Catalogue Executive Summary, http://pegasus.phast.umass.edu/
\end{thebibliography}
\end{document}